\documentclass[showpacs,preprintnumbers,prl,twocolumn]{revtex4}
\usepackage{amssymb}
\usepackage{amsmath}
\usepackage{graphicx}
\usepackage{bm}
\usepackage{amsfonts}

\newcommand{\ket}[1]{\mbox{$|#1\rangle$}}

\def\be{\begin{equation}} 
\def\ee{\end{equation}}

\begin{document}

\title{High coherence hybrid superconducting qubit}

\author{Matthias Steffen, Shwetank Kumar, David P. DiVincenzo, J.R. Rozen, George A. Keefe, Mary Beth Rothwell, Mark B. Ketchen}

\affiliation{{IBM Watson Research Ctr., Yorktown Heights, NY 10598
USA}}

\begin{abstract}
We measure the coherence of a new superconducting qubit, the {\em low-impedance flux qubit}, finding $T_2^* \sim T_1 \sim 1.5\mu$s. It is a three-junction flux qubit, but the ratio of junction critical currents is chosen to make the qubit's potential have a single well form.  The low impedance of its large shunting capacitance protects it from decoherence.  This qubit has a moderate anharmonicity, whose sign is reversed compared with all other popular qubit designs.  The qubit is capacitively coupled to a high-Q resonator in a $\lambda/2$ configuration, which permits the qubit's state to be read out dispersively.

\end{abstract}

\maketitle

While there have been many successful superconducting qubit types, their large diversity suggests that the optimal qubit will be a hybrid combining favorable features of all: the tunability of the flux qubit \cite{Mooij99,Chiorescu03,Niskanen06}, the simplicity, robustness and low impedance of the phase qubit \cite{Ansmann09,Hofheinz09,Steffen06b} and the high coherence and compatibility with high-Q superconducting resonators of the transmon \cite{DiCarlo09,Koch07}. We have built such a hybrid, related to a suggested design of You {\em et al.}\cite{You07}.  Our capacitively shunted flux qubit begins as a traditional three-junction loop\cite{Mooij99}, but is made to have low impedance by virtue of a large capacitive shunt ($C_s=100$fF) of the small junction.  This new superconducting qubit is as coherent as the best currently reported; we measure $T_2^* \sim T_1 \sim 1.5\mu$s.

Since the key to this qubit is the large shunting capacitance $C_s$ and therefore its low effective impedance $\sqrt{L_J/C_s}$, we will call it the {\em low-impedance flux qubit} ($_Z$flux qubit).  As Fig. 1(a) shows, the shunt capacitor is realized using a simple, reliable single-level interdigitated structure.  We choose the ratio of the small and large junction critical currents $I_0$ to be around $\alpha=0.3$.  For this $\alpha$ the qubit potential has only one minimum (see Eq. (3) below), and the qubit shows only a weak dependence of the qubit frequency $\omega_{01}$ on applied flux $\Phi$. As for the original flux qubit, a ``sweet spot" exists at which the qubit is to first order insensitive to $\Phi$, giving rise to long dephasing times, but even away from this degeneracy point our frequency sensitivity is about a factor of $30$ smaller than in the traditional flux qubit. Our flux sensitivity is comparable to that of the phase qubit ($\partial\omega_{01}/\partial\Phi\sim 30$ GHz/$\Phi_0$) which permits tunability without completely destroying phase coherence, despite the presence of significant flux noise amplitude on the order of $S_{\Phi}=1-2\mu \Phi_0/\sqrt{\mathrm{Hz}}$. Modeling indicates that our qubit at the sweet spot still has appreciable anharmonicity, with $|\omega_{12}-\omega_{01}|/2 \pi$ in the neighborhood of several 100 MHz (or about $2-10\%$ of the qubit resonance frequency, depending on $\alpha$), but interestingly, with $\omega_{12}>\omega_{01}$, the opposite of any important qubit except the flux qubit. Such anharmonicity leads to a situation where all of the lowest energy levels for a two-qubit system would be those of the computational manifold $\ket{0}$ and $\ket{1}$, which will facilitate coupled qubit experiments. 

\begin{figure}[htp]
	\centering
		\includegraphics[width=0.5\textwidth]{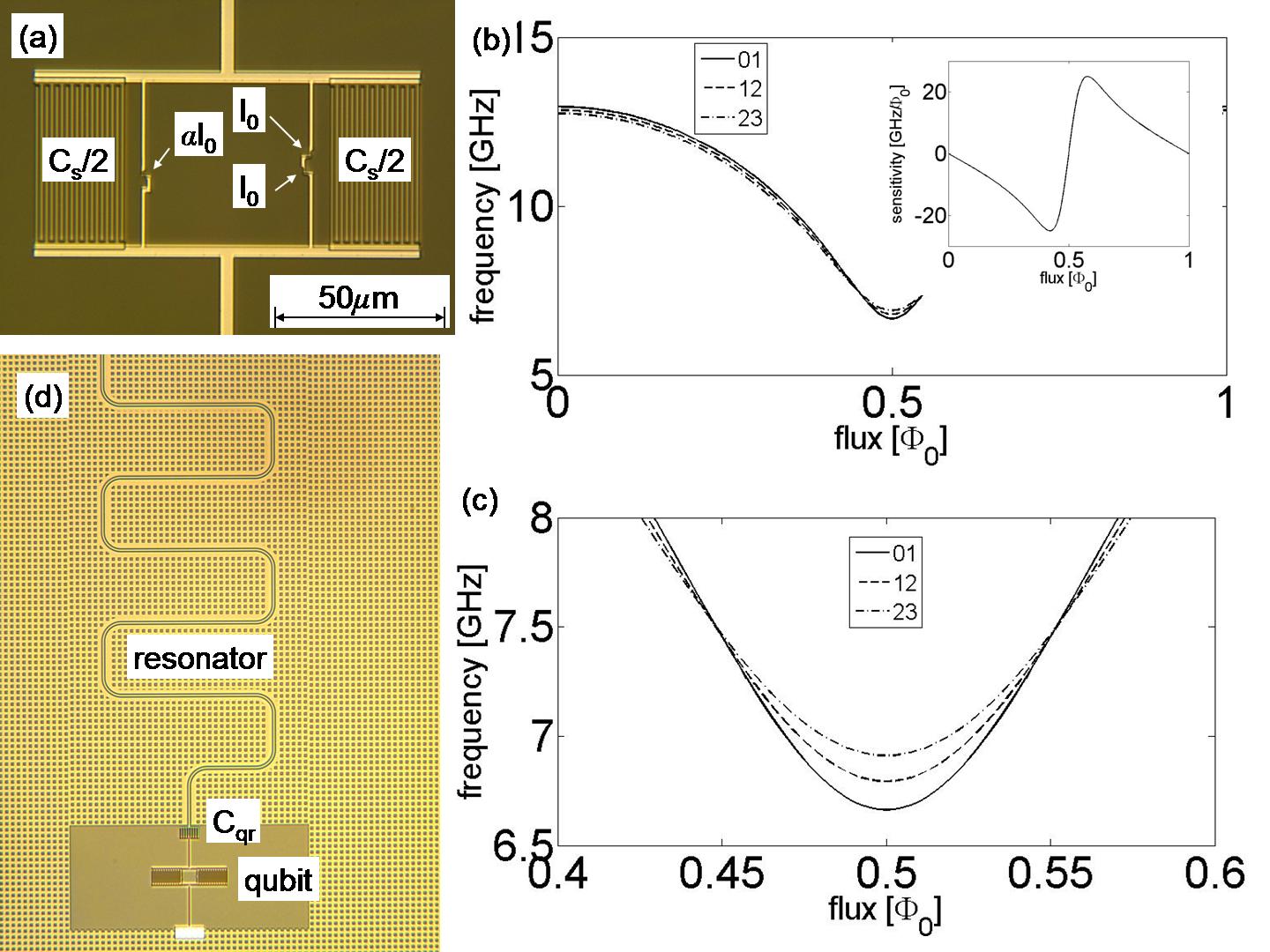}
	\caption{Micrograph and simulated frequency response of the $_Z$flux qubit. (a) A micrograph of the qubit shows the interdigitated shunting capacitor ($C_s=100$fF), which is made of aluminum simultaneously with the junction fabrication step. The qubit has three junctions, as in the traditional flux qubit. (b) The qubit frequency response is much weaker than the traditional flux qubit (inset: derivative $\partial f_{01}/\partial \Phi_0$). Near the sweet spot at $\Phi=0.5\Phi_0$ the qubit anharmonicity is ``inverted": $\omega_{12}>\omega_{01}$. (d) The qubit is read out dispersively by coupling it capacitively via $C_{qr} \sim 1.6fF$ to a half-wavelength coplanar waveguide resonator with a resonance frequency of $f_r=10.35$GHz.}
	\label{fig:fig1}
\end{figure}

The reduced impedance of this qubit has several advantages. Qubits with low self-capacitance are more susceptible to residual capacitive coupling effects \cite{Steffen09a,Paauw08}. By increasing the self-capacitance to a level similar to the phase qubit and the transmon, these effects are still present but at a more manageable level. Additionally, by introducing a large self-capacitance we provide additional means of coupling multiple qubits effectively (capacitive or inductive coupling).

The $_Z$flux qubit is modeled as in reference \cite{You07}. Assuming the loop inductance is much less than the junction inductance $L \ll L_J$ the potential energy is two-dimensional; in terms of sum and difference phases $\delta_{p,m}$,

\begin{equation}
U_{2D} = -2E_J\mathrm{cos}(\delta_p/2)\mathrm{cos}(\delta_m/2)-\alpha E_J \mathrm{cos}(2\pi\Phi/\Phi_0 - \delta_m)
\label{eq:2dpot}
\end{equation}

here $E_J=I_0\Phi_0/2\pi$ and $\Phi_0=h/2e$. The kinetic term is:

\begin{equation}
K_{2D} = \left( \frac{\partial}{\partial \delta_p} \right)^2 \frac{4e^2}{C_J} + \left( \frac{\partial}{\partial \delta_m} \right)^2 \frac{4 e^2}{C_J(1+2\beta)}
\label{eq:2dkin}
\end{equation}

where $\beta=\alpha+C_s/C_J$.  $C_J$ is the capacitance of the larger Josephson junctions (typical value is $\sim 5-10$fF for shadow evaporated junctions); we assume that the capacitance of the small junction is $\alpha C_J$.  The introduction of a large shunting capacitance $C_s>>C_J$ permits the $\delta_p$ direction to be safely ignored, as confirmed by a Born-Oppenheimer analysis \cite{DiVincenzo06}. Thus the potential is accurately represented in a simple one-dimensional form:
\begin{equation}
U_{1D} = -2E_J\mathrm{cos}(\delta/2)- \alpha E_J\mathrm{cos}(2\pi\Phi/\Phi_0 - \delta).
\label{eq:1dpot}
\end{equation}

Only the second kinetic term remains from eq. \ref{eq:2dkin} with an effective capacitance approximately equal to $C_s$. The expected frequency dependence of the first four energy levels is shown in Fig. \ref{fig:fig1}(b) and (c) as well as the derivative (inset). We note that because the potential has only one minimum, it is not practical to measure the state of the $_Z$flux qubit using a magnetometer -- but it is quite practical to probe it dispersively via a cavity \cite{DiCarlo09,Steffen10b}. 

\begin{figure}[htp]
	\centering
		\includegraphics[width=0.5\textwidth]{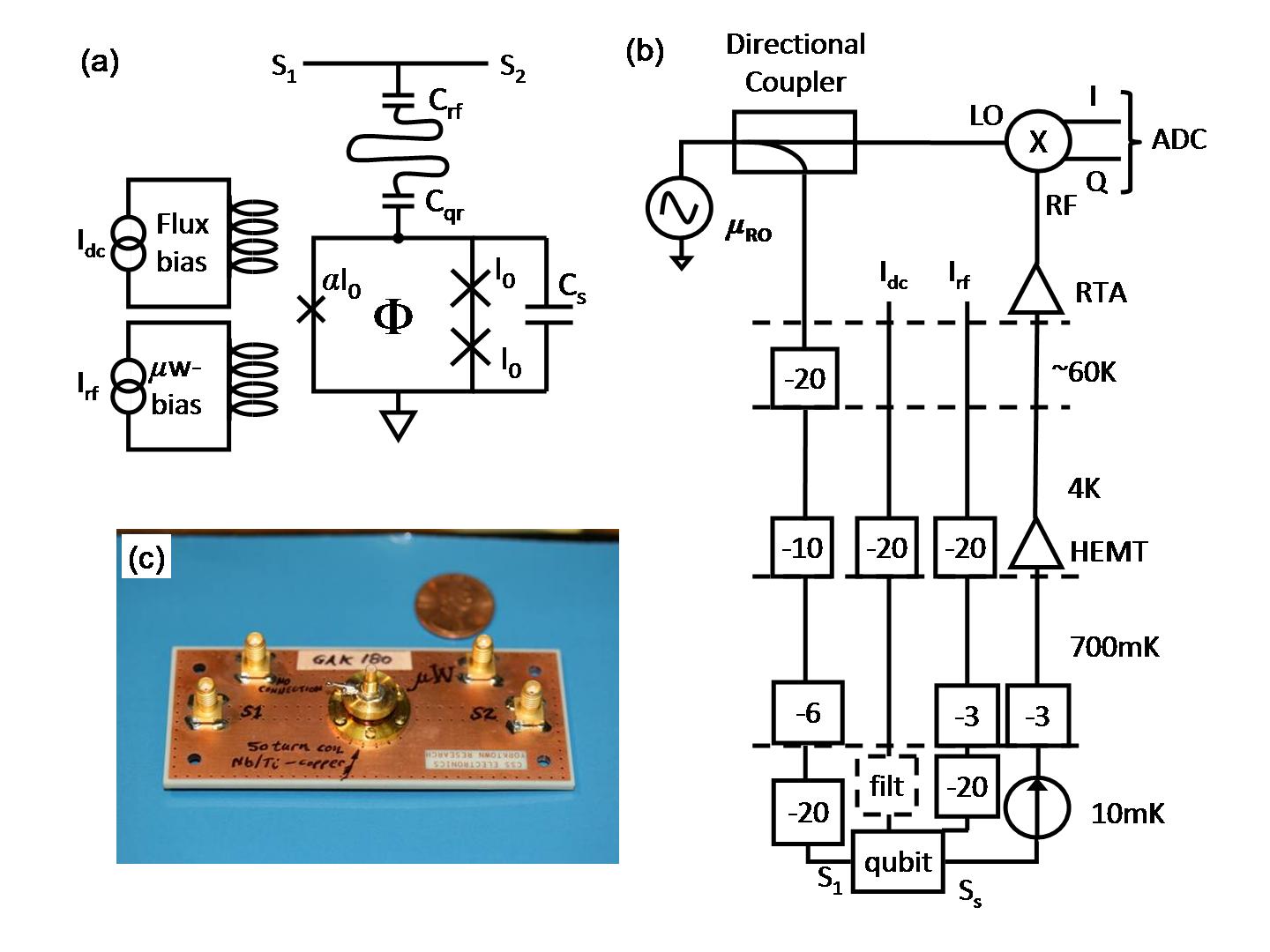}
	\caption{Experimental setup. (a) The resonator is capacitively coupled to a feedline. Both the microwaves and dc bias are inductively coupled to the qubit. The junction self-capcitance is not included in the drawing. (b) The qubit is measured dispersively using standard microwave techniques using an IQ-mixer setup. (c) The qubit is mounted on a copper pc board with SSMA connectors. The dc flux bias is generated using a hand-wound external superconducting wire coil mounted on the lid of the box.}
	\label{fig:fig2}
\end{figure}

We fabricated a $_Z$flux qubit (closely following steps outlined in \cite{Steffen10}) on a 200mm high resistivity ($> 1000 \Omega cm$) silicon (Si) wafer without any thermal oxide. The feed line and resonator as well as the corresponding ground plane are made of 200nm thick niobium and deposited by physical vapor deposition (PVD). The patterning is done using deep UV lithography followed by a reactive ion etch (RIE) in a chlorine based plasma. Intrinsic quality factors of resonators made separately ($10 \mu$m center strip width, $6 \mu$m gap) are measured to be $Q=80k-100k$, confirming clean substrates, and confirming that processing does not introduce significant defects. The Josephson junctions require a second mask, which is created out of a LOR5A/Ge/PMMA trilayer which is patterned by ebeam lithography, then developed, and then etched using a CF$_4$/Ar plasma. The bottom layer is wet etched using OPD7262. The substrate surface is pre-cleaned using an ion mill, followed by depositing two aluminum layers, separated by a brief oxidation of the first layer. The shunting capacitor is formed during the Al deposition with $2 \mu$m lines and spaces.

\begin{figure}[htp]
	\centering
		\includegraphics[width=0.5\textwidth]{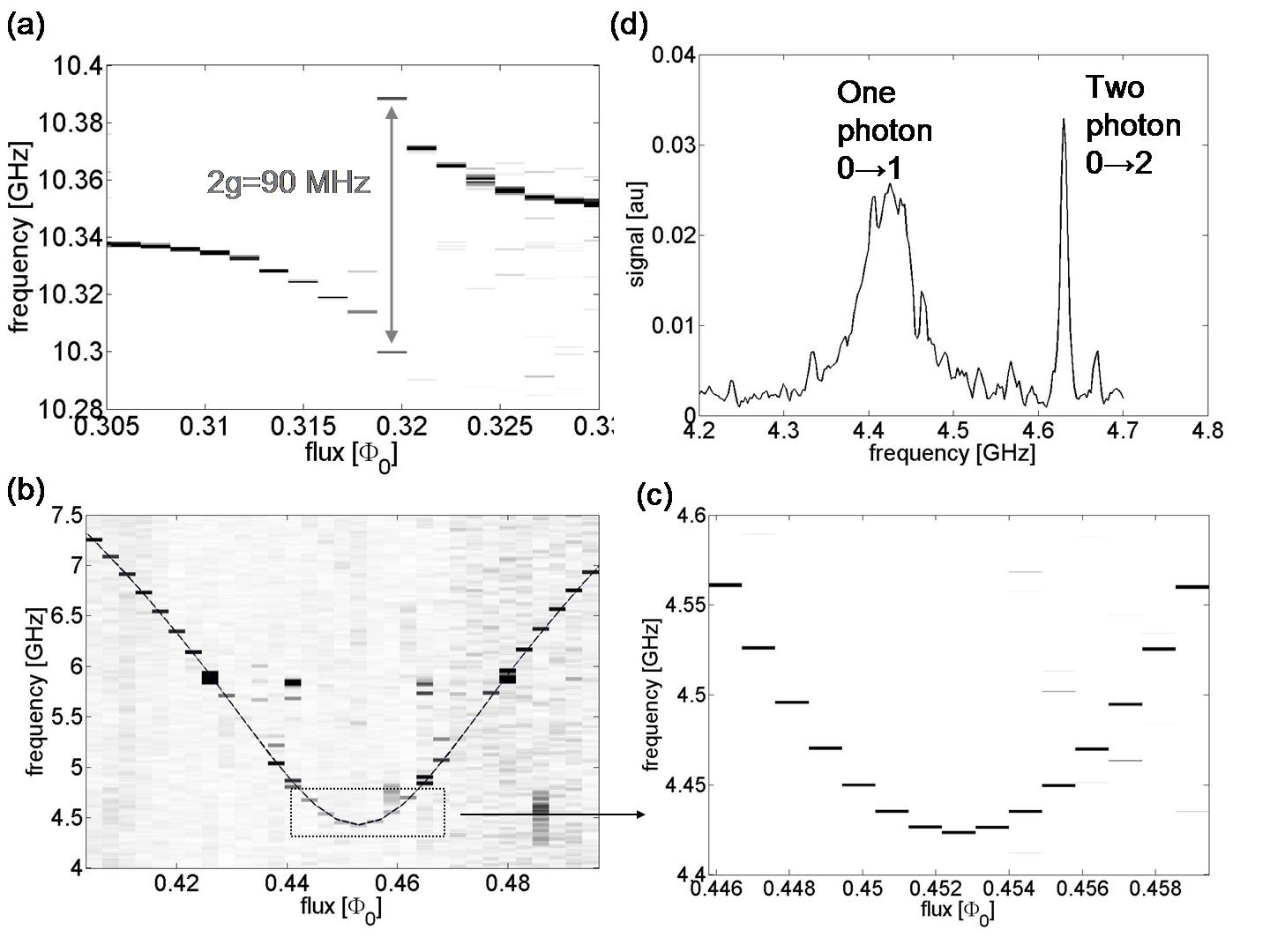}
	\caption{Qubit Spectroscopy. (a) The vacuum-Rabi splitting indicates $g=45$MHz. (b) Qubit spectroscopy over a broad flux range agrees well with the one-dimensional theory (dashed line). The data is significantly broadened in software for the qubit spectroscopy to be visible on this scale. A box mode near $5.6$GHz couples microwaves much more strongly to the qubit than all other frequencies. (c) Qubit spectroscopy near the sweet spot (data not broadened) reminds us of the typical parabolic response of the flux qubit. (d) The two-photon $\ket{0} \rightarrow \ket{2}$ transition is higher in frequency, revealing the ``inverted" anharmonicity of the $_Z$flux qubit.}
	\label{fig:fig3}
\end{figure}

We implemented the required dispersive read-out techniques and measured the qubit; as Fig. \ref{fig:fig1}d indicates, the qubit is grounded and end-coupled via a coupling capacitor (C$_{qr} \sim 1.5$fF) to a standard coplanar waveguide resonator, which is in turn capacitively coupled ($C_{rf} \sim 2$fF) to a feed line, similar to \cite{Steffen10b}. We first measured the vacuum-Rabi splitting (Fig. \ref{fig:fig3}a) and obtain $g=45$MHz, consistent with the magnitude of the small coupling capacitance $C_{qr}$. We note that the cavity quality factor $Q \sim 60,000$ is quite large and increases the cavity's response time, impacting the signal-to-noise ratio in our experiment \cite{Blais04}. The qubit spectroscopy over a broad flux range is shown in Fig. \ref{fig:fig3}(b) and agrees well with the predicted frequencies from Eq. (\ref{eq:1dpot}) using $I_0=0.34\mu$A, $\alpha=0.43$, and $C_s=110$fF (dashed line). Zooming in on the sweet spot (Fig. \ref{fig:fig3}(c)) reveals a flux insensitive qubit response corresponding to $\omega_{10}/2\pi = 4.4225$GHz. At the sweet spot the anharmonicity of the qubit is measured by applying a large microwave drive and observing the two-photon $\ket{0} \rightarrow \ket{2}$ transition, which is higher in frequency than $\omega_{10}/2\pi$, clearly showing the inverted anharmonicity. Separate experiments confirm this by observing Rabi oscillations between the $\ket{1}$ and $\ket{2}$ states, and by identifying a {\em negative} dispersive shift of the resonator frequency (data not shown). The detuning $(\omega_{12}-\omega_{10})/2\pi \sim 400$MHz is consistent with parameters obtained from fitting $\omega_{10}$ above.

\begin{figure}[htp]
	\centering
		\includegraphics[width=0.5\textwidth]{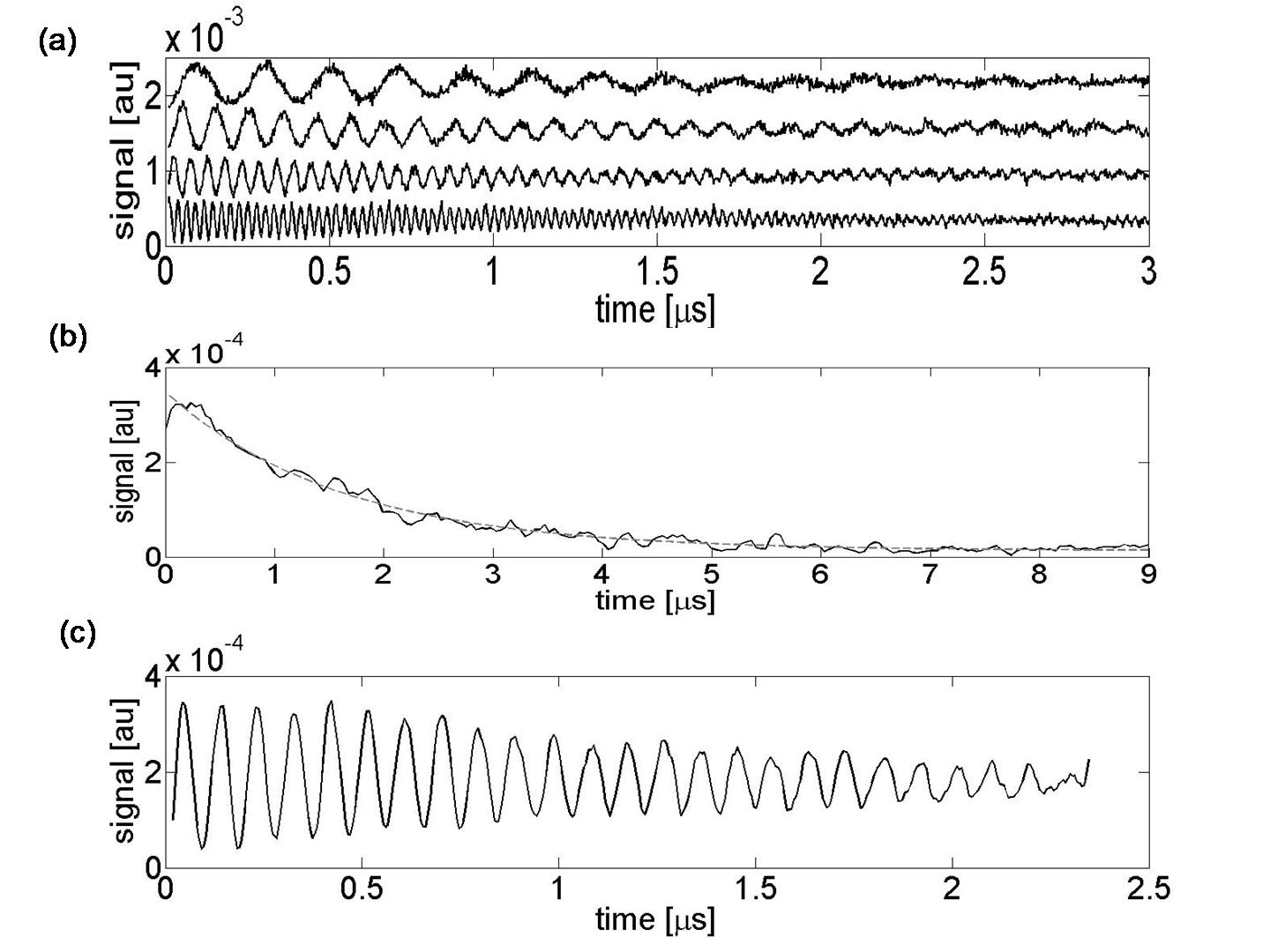}
	\caption{Coherence time measurements of the qubit. (a) Rabi oscillations for varying microwave amplitudes (factor 2x difference for each trace). (b) A direct measure of the energy decay is obtained from a fit to the data (grey dashed line) and is found to be $T_1=1.6 \mu$s. (c) Ramsey fringes are implemented by applying a Hadamard gate ($180^\circ$ degree rotation around an axis tilted by $45^\circ$ from the horizontal axis). The detuning between the qubit and microwave frequency is $\sim 10$MHz, consistent with the observed period. The data can be fitted to an exponential decay with a decay constant $T_2^* \sim 1.3 \mu$s.}
	\label{fig:fig4}
\end{figure}

Our $_Z$flux qubit has remarkably long coherence times. Fig. \ref{fig:fig4}(a) shows a Rabi trace, for various microwave drive amplitudes, indicating a decay constant of nearly 2$\mu$s.  A direct measurement of $T_1$ (Fig. \ref{fig:fig4}(c)), which detects the return of the qubit to the ground state after $\pi$ pulsing, indicates $T_1=1.6\mu$s, corresponding to $Q=T_1\omega_{10}=44k$. Fig. \ref{fig:fig4}(c) shows a Ramsey spectrum, showing $T_2^*=1.3\mu$s.  Refocussing pulses are not used here, but the DC flux is recalibrated back to the sweet spot every few minutes during the experiment to correct for drift. Finally, spin echo data indicates $T_2=1.5\mu$s (data not included). The fact that $T_2 \neq 2T_1$ indicates dephasing noise is not due to low- but rather to high-frequency fluctuations, possibly in the kHz-MHz regime. Further repeatability studies must be conducted to determine if such noise is consistently present.

The energy decay for a higher qubit frequency $\omega_{10}/2\pi=7.12$GHz is found to be $T_1=0.86 \mu$s, corresponding to $Q=38.5k$, not too disimilar from the measured value at the sweet spot. Disspiation from dielectric loss predicts a frequency independent quality factor, and hence we believe that the energy decay is limited by dielectric loss, presumably by the native oxide formed on the aluminum of the interdigitated capacitor. Although Q appears to vary slightly, we have not performed an exhaustive search to extract Q over all frequencies. $Q=40k$ is close to the reported value for the transmon (Q near $70k$ \cite{Houck09}), which is also suspected to be set by dielectric loss.

As for all other superconducting qubits, the $_Z$flux qubit permits the anharmonicity, resonance frequency and flux sensitivity to be traded off against each other. Such a trade-off optimization is particularly interesting when introducing a tunable ratio $\alpha$ of the critical currents by replacing the small junction with a SQUID \cite{Steffen10,Paauw08}. Although such an optimization is not the goal in this first demonstration, we believe that the $_Z$flux qubit can be made to retain the inverted anharmonicity yet still be tunable with a very weak dependency of the frequency on flux. The rich physics allowed by introducing additional junctions together with the increased self-capacitance is a promising avenue for designing highly scalable qubits.

Several features are clearly to be improved in further experiments.  The coupling to the qubit here is set to be very weak, so that the dispersive shift used in the detection is very small ($\sim$40kHz) and the qubit state cannot be acquired in a single shot. Modest increases of the coupling capacitances should permit much reduced averaging times, and possibly single-shot read out.  Further spectroscopy studies, with careful power-dependence studies, should permit a direct identification of other important states, particularly in the regime where some of the higher lying energy levels are degenerate.  Still, the present experiments already tell us a wealth of new information about capacitively shunted qubits. Feared decoherence mechanisms such as dielectric loss, two-level systems, and quasiparticle dissipation still permit coherence times at the 1$\mu$s level.

The views and conclusions contained in this document are those of the authors and should not be interpreted as representing the official policies, either expressly or implied, of the U.S. Government.

\bibliography{bibmaster}

\begin{thebibliography}{16}
\expandafter\ifx\csname natexlab\endcsname\relax\def\natexlab#1{#1}\fi
\expandafter\ifx\csname bibnamefont\endcsname\relax
  \def\bibnamefont#1{#1}\fi
\expandafter\ifx\csname bibfnamefont\endcsname\relax
  \def\bibfnamefont#1{#1}\fi
\expandafter\ifx\csname citenamefont\endcsname\relax
  \def\citenamefont#1{#1}\fi
\expandafter\ifx\csname url\endcsname\relax
  \def\url#1{\texttt{#1}}\fi
\expandafter\ifx\csname urlprefix\endcsname\relax\def\urlprefix{URL }\fi
\providecommand{\bibinfo}[2]{#2}
\providecommand{\eprint}[2][]{\url{#2}}

\bibitem[{\citenamefont{Mooij et~al.}(1999)\citenamefont{Mooij, Orlando,
  Levitov, Tian, van~der Wal, and Lloyd}}]{Mooij99}
\bibinfo{author}{\bibfnamefont{J.}~\bibnamefont{Mooij}},
  \bibinfo{author}{\bibfnamefont{T.}~\bibnamefont{Orlando}},
  \bibinfo{author}{\bibfnamefont{L.}~\bibnamefont{Levitov}},
  \bibinfo{author}{\bibfnamefont{L.}~\bibnamefont{Tian}},
  \bibinfo{author}{\bibfnamefont{C.~H.} \bibnamefont{van~der Wal}},
  \bibnamefont{and} \bibinfo{author}{\bibfnamefont{S.}~\bibnamefont{Lloyd}},
  \bibinfo{journal}{Science} \textbf{\bibinfo{volume}{285}},
  \bibinfo{pages}{1036 } (\bibinfo{year}{1999}).

\bibitem[{\citenamefont{Chiorescu et~al.}(2003)\citenamefont{Chiorescu,
  Nakamura, Harmans, and Mooij}}]{Chiorescu03}
\bibinfo{author}{\bibfnamefont{I.}~\bibnamefont{Chiorescu}},
  \bibinfo{author}{\bibfnamefont{Y.}~\bibnamefont{Nakamura}},
  \bibinfo{author}{\bibfnamefont{C.}~\bibnamefont{Harmans}}, \bibnamefont{and}
  \bibinfo{author}{\bibfnamefont{J.}~\bibnamefont{Mooij}},
  \bibinfo{journal}{Science} \textbf{\bibinfo{volume}{299}},
  \bibinfo{pages}{1869} (\bibinfo{year}{2003}).

\bibitem[{\citenamefont{Niskanen et~al.}(2006)\citenamefont{Niskanen, Harrabi,
  Yoshihara, Nakamura, and Tsai}}]{Niskanen06}
\bibinfo{author}{\bibfnamefont{A.~O.} \bibnamefont{Niskanen}},
  \bibinfo{author}{\bibfnamefont{K.}~\bibnamefont{Harrabi}},
  \bibinfo{author}{\bibfnamefont{F.}~\bibnamefont{Yoshihara}},
  \bibinfo{author}{\bibfnamefont{Y.}~\bibnamefont{Nakamura}}, \bibnamefont{and}
  \bibinfo{author}{\bibfnamefont{J.~S.} \bibnamefont{Tsai}},
  \bibinfo{journal}{Phys. Rev. B} \textbf{\bibinfo{volume}{74}},
  \bibinfo{pages}{220503} (\bibinfo{year}{2006}).

\bibitem[{\citenamefont{Ansmann et~al.}(2009)\citenamefont{Ansmann, Wang,
  Bialczak, Hofheinz, Lucero, Neeley, O'Connell, Sank, Weides, Wenner
  et~al.}}]{Ansmann09}
\bibinfo{author}{\bibfnamefont{M.}~\bibnamefont{Ansmann}},
  \bibinfo{author}{\bibfnamefont{H.}~\bibnamefont{Wang}},
  \bibinfo{author}{\bibfnamefont{R.}~\bibnamefont{Bialczak}},
  \bibinfo{author}{\bibfnamefont{M.}~\bibnamefont{Hofheinz}},
  \bibinfo{author}{\bibfnamefont{E.}~\bibnamefont{Lucero}},
  \bibinfo{author}{\bibfnamefont{M.}~\bibnamefont{Neeley}},
  \bibinfo{author}{\bibfnamefont{A.}~\bibnamefont{O'Connell}},
  \bibinfo{author}{\bibfnamefont{D.}~\bibnamefont{Sank}},
  \bibinfo{author}{\bibfnamefont{M.}~\bibnamefont{Weides}},
  \bibinfo{author}{\bibfnamefont{J.}~\bibnamefont{Wenner}},
  \bibnamefont{et~al.}, \bibinfo{journal}{Nature}
  \textbf{\bibinfo{volume}{461}}, \bibinfo{pages}{504} (\bibinfo{year}{2009}).

\bibitem[{\citenamefont{Hofheinz et~al.}(2009)\citenamefont{Hofheinz, Wang,
  Ansmann, Bialczak, Lucero, Neeley, O'Connell, Sank, Martinis, and
  Cleland}}]{Hofheinz09}
\bibinfo{author}{\bibfnamefont{M.}~\bibnamefont{Hofheinz}},
  \bibinfo{author}{\bibfnamefont{H.}~\bibnamefont{Wang}},
  \bibinfo{author}{\bibfnamefont{M.}~\bibnamefont{Ansmann}},
  \bibinfo{author}{\bibfnamefont{R.}~\bibnamefont{Bialczak}},
  \bibinfo{author}{\bibfnamefont{E.}~\bibnamefont{Lucero}},
  \bibinfo{author}{\bibfnamefont{M.}~\bibnamefont{Neeley}},
  \bibinfo{author}{\bibfnamefont{A.}~\bibnamefont{O'Connell}},
  \bibinfo{author}{\bibfnamefont{D.~W.~J.} \bibnamefont{Sank}},
  \bibinfo{author}{\bibfnamefont{J.}~\bibnamefont{Martinis}}, \bibnamefont{and}
  \bibinfo{author}{\bibfnamefont{A.}~\bibnamefont{Cleland}},
  \bibinfo{journal}{Nature} \textbf{\bibinfo{volume}{459}},
  \bibinfo{pages}{546} (\bibinfo{year}{2009}).

\bibitem[{\citenamefont{Steffen et~al.}(2006)\citenamefont{Steffen, Ansmann,
  Bialczak, Katz, Lucero, McDermott, Neeley, Weig, Cleland, and
  Martinis}}]{Steffen06b}
\bibinfo{author}{\bibfnamefont{M.}~\bibnamefont{Steffen}},
  \bibinfo{author}{\bibfnamefont{M.}~\bibnamefont{Ansmann}},
  \bibinfo{author}{\bibfnamefont{R.}~\bibnamefont{Bialczak}},
  \bibinfo{author}{\bibfnamefont{N.}~\bibnamefont{Katz}},
  \bibinfo{author}{\bibfnamefont{E.}~\bibnamefont{Lucero}},
  \bibinfo{author}{\bibfnamefont{R.}~\bibnamefont{McDermott}},
  \bibinfo{author}{\bibfnamefont{M.}~\bibnamefont{Neeley}},
  \bibinfo{author}{\bibfnamefont{E.}~\bibnamefont{Weig}},
  \bibinfo{author}{\bibfnamefont{A.}~\bibnamefont{Cleland}}, \bibnamefont{and}
  \bibinfo{author}{\bibfnamefont{J.}~\bibnamefont{Martinis}},
  \bibinfo{journal}{Science} \textbf{\bibinfo{volume}{313}},
  \bibinfo{pages}{1423} (\bibinfo{year}{2006}).

\bibitem[{\citenamefont{DiCarlo et~al.}(2009)\citenamefont{DiCarlo, Chow,
  Bambetta, Bishop, Schuster, Majer, Blais, Frunzio, Girvin, and
  Schoelkopf}}]{DiCarlo09}
\bibinfo{author}{\bibfnamefont{L.}~\bibnamefont{DiCarlo}},
  \bibinfo{author}{\bibfnamefont{J.}~\bibnamefont{Chow}},
  \bibinfo{author}{\bibfnamefont{J.}~\bibnamefont{Bambetta}},
  \bibinfo{author}{\bibfnamefont{L.~S.} \bibnamefont{Bishop}},
  \bibinfo{author}{\bibfnamefont{D.}~\bibnamefont{Schuster}},
  \bibinfo{author}{\bibfnamefont{J.}~\bibnamefont{Majer}},
  \bibinfo{author}{\bibfnamefont{A.}~\bibnamefont{Blais}},
  \bibinfo{author}{\bibfnamefont{L.}~\bibnamefont{Frunzio}},
  \bibinfo{author}{\bibfnamefont{S.}~\bibnamefont{Girvin}}, \bibnamefont{and}
  \bibinfo{author}{\bibfnamefont{R.}~\bibnamefont{Schoelkopf}},
  \bibinfo{journal}{Nature} \textbf{\bibinfo{volume}{460}},
  \bibinfo{pages}{240} (\bibinfo{year}{2009}).

\bibitem[{\citenamefont{Koch et~al.}(2007)\citenamefont{Koch, Yu, Gambetta,
  Houck, Schuster, Majer, Blais, Devoret, Girvin, and Schoelkopf}}]{Koch07}
\bibinfo{author}{\bibfnamefont{J.}~\bibnamefont{Koch}},
  \bibinfo{author}{\bibfnamefont{T.~M.} \bibnamefont{Yu}},
  \bibinfo{author}{\bibfnamefont{J.}~\bibnamefont{Gambetta}},
  \bibinfo{author}{\bibfnamefont{A.~A.} \bibnamefont{Houck}},
  \bibinfo{author}{\bibfnamefont{D.~I.} \bibnamefont{Schuster}},
  \bibinfo{author}{\bibfnamefont{J.}~\bibnamefont{Majer}},
  \bibinfo{author}{\bibfnamefont{A.}~\bibnamefont{Blais}},
  \bibinfo{author}{\bibfnamefont{M.~H.} \bibnamefont{Devoret}},
  \bibinfo{author}{\bibfnamefont{S.~M.} \bibnamefont{Girvin}},
  \bibnamefont{and} \bibinfo{author}{\bibfnamefont{R.~J.}
  \bibnamefont{Schoelkopf}}, \bibinfo{journal}{Phys. Rev. A}
  \textbf{\bibinfo{volume}{76}}, \bibinfo{pages}{042319}
  (\bibinfo{year}{2007}).

\bibitem[{\citenamefont{You et~al.}(2007)\citenamefont{You, Hu, Ashhab, and
  Nori}}]{You07}
\bibinfo{author}{\bibfnamefont{J.}~\bibnamefont{You}},
  \bibinfo{author}{\bibfnamefont{X.}~\bibnamefont{Hu}},
  \bibinfo{author}{\bibfnamefont{S.}~\bibnamefont{Ashhab}}, \bibnamefont{and}
  \bibinfo{author}{\bibfnamefont{F.}~\bibnamefont{Nori}},
  \bibinfo{journal}{Phys. Rev. B} \textbf{\bibinfo{volume}{75}},
  \bibinfo{pages}{140515} (\bibinfo{year}{2007}).

\bibitem[{\citenamefont{Steffen et~al.}(2009)\citenamefont{Steffen, Brito,
  DiVincenzo, Kumar, and Ketchen}}]{Steffen09a}
\bibinfo{author}{\bibfnamefont{M.}~\bibnamefont{Steffen}},
  \bibinfo{author}{\bibfnamefont{F.}~\bibnamefont{Brito}},
  \bibinfo{author}{\bibfnamefont{D.}~\bibnamefont{DiVincenzo}},
  \bibinfo{author}{\bibfnamefont{S.}~\bibnamefont{Kumar}}, \bibnamefont{and}
  \bibinfo{author}{\bibfnamefont{M.}~\bibnamefont{Ketchen}},
  \bibinfo{journal}{New J. Phys.} \textbf{\bibinfo{volume}{11}},
  \bibinfo{pages}{033030} (\bibinfo{year}{2009}).

\bibitem[{\citenamefont{Paauw et~al.}(2009)\citenamefont{Paauw, Fedorov,
  Harmans, and Mooij}}]{Paauw08}
\bibinfo{author}{\bibfnamefont{F.}~\bibnamefont{Paauw}},
  \bibinfo{author}{\bibfnamefont{A.}~\bibnamefont{Fedorov}},
  \bibinfo{author}{\bibfnamefont{C.}~\bibnamefont{Harmans}}, \bibnamefont{and}
  \bibinfo{author}{\bibfnamefont{J.}~\bibnamefont{Mooij}},
  \bibinfo{journal}{Phys. Rev. Lett.} \textbf{\bibinfo{volume}{102}},
  \bibinfo{pages}{090501} (\bibinfo{year}{2009}).

\bibitem[{\citenamefont{DiVincenzo et~al.}(2006)\citenamefont{DiVincenzo,
  Brito, and Koch}}]{DiVincenzo06}
\bibinfo{author}{\bibfnamefont{D.~P.} \bibnamefont{DiVincenzo}},
  \bibinfo{author}{\bibfnamefont{F.}~\bibnamefont{Brito}}, \bibnamefont{and}
  \bibinfo{author}{\bibfnamefont{R.~H.} \bibnamefont{Koch}},
  \bibinfo{journal}{Phys. Rev. B} \textbf{\bibinfo{volume}{74}},
  \bibinfo{pages}{014514} (\bibinfo{year}{2006}).

\bibitem[{\citenamefont{Steffen
  et~al.}(2010{\natexlab{a}})\citenamefont{Steffen, Kumar, ViVincenzo, Keefe,
  Ketchen, Rothwell, and Rozen}}]{Steffen10b}
\bibinfo{author}{\bibfnamefont{M.}~\bibnamefont{Steffen}},
  \bibinfo{author}{\bibfnamefont{S.}~\bibnamefont{Kumar}},
  \bibinfo{author}{\bibfnamefont{D.~D.} \bibnamefont{ViVincenzo}},
  \bibinfo{author}{\bibfnamefont{G.}~\bibnamefont{Keefe}},
  \bibinfo{author}{\bibfnamefont{M.}~\bibnamefont{Ketchen}},
  \bibinfo{author}{\bibfnamefont{M.~B.} \bibnamefont{Rothwell}},
  \bibnamefont{and} \bibinfo{author}{\bibfnamefont{J.}~\bibnamefont{Rozen}},
  \bibinfo{journal}{Appl. Phys. Lett.} \textbf{\bibinfo{volume}{96}},
  \bibinfo{pages}{102506} (\bibinfo{year}{2010}{\natexlab{a}}).

\bibitem[{\citenamefont{Steffen
  et~al.}(2010{\natexlab{b}})\citenamefont{Steffen, Brito, DiVincenzo,
  Farinelli, Keefe, Ketchen, Kumar, Milliken, Rothwell, Rozen
  et~al.}}]{Steffen10}
\bibinfo{author}{\bibfnamefont{M.}~\bibnamefont{Steffen}},
  \bibinfo{author}{\bibfnamefont{F.}~\bibnamefont{Brito}},
  \bibinfo{author}{\bibfnamefont{D.~P.} \bibnamefont{DiVincenzo}},
  \bibinfo{author}{\bibfnamefont{M.}~\bibnamefont{Farinelli}},
  \bibinfo{author}{\bibfnamefont{G.}~\bibnamefont{Keefe}},
  \bibinfo{author}{\bibfnamefont{M.}~\bibnamefont{Ketchen}},
  \bibinfo{author}{\bibfnamefont{S.}~\bibnamefont{Kumar}},
  \bibinfo{author}{\bibfnamefont{F.~P.} \bibnamefont{Milliken}},
  \bibinfo{author}{\bibfnamefont{M.~B.} \bibnamefont{Rothwell}},
  \bibinfo{author}{\bibfnamefont{J.}~\bibnamefont{Rozen}},
  \bibnamefont{et~al.}, \bibinfo{journal}{Journal of Physics: Condensed Matter}
  \textbf{\bibinfo{volume}{22}}, \bibinfo{pages}{053201}
  (\bibinfo{year}{2010}{\natexlab{b}}).

\bibitem[{\citenamefont{Blais et~al.}(2004)\citenamefont{Blais, Huang,
  Wallraff, Girvin, and Schoelkopf}}]{Blais04}
\bibinfo{author}{\bibfnamefont{P.}~\bibnamefont{Blais}},
  \bibinfo{author}{\bibfnamefont{R.-S.} \bibnamefont{Huang}},
  \bibinfo{author}{\bibfnamefont{A.}~\bibnamefont{Wallraff}},
  \bibinfo{author}{\bibfnamefont{S.}~\bibnamefont{Girvin}}, \bibnamefont{and}
  \bibinfo{author}{\bibfnamefont{R.}~\bibnamefont{Schoelkopf}},
  \bibinfo{journal}{Phys. Rev. A} \textbf{\bibinfo{volume}{69}},
  \bibinfo{pages}{062320} (\bibinfo{year}{2004}).

\bibitem[{\citenamefont{Houck et~al.}(2009)\citenamefont{Houck, Koch, Devoret,
  Girvin, and Schoelkopf}}]{Houck09}
\bibinfo{author}{\bibfnamefont{A.}~\bibnamefont{Houck}},
  \bibinfo{author}{\bibfnamefont{J.}~\bibnamefont{Koch}},
  \bibinfo{author}{\bibfnamefont{M.}~\bibnamefont{Devoret}},
  \bibinfo{author}{\bibfnamefont{S.}~\bibnamefont{Girvin}}, \bibnamefont{and}
  \bibinfo{author}{\bibfnamefont{R.}~\bibnamefont{Schoelkopf}},
  \bibinfo{journal}{Qunt. Inf. Proc.} \textbf{\bibinfo{volume}{8}},
  \bibinfo{pages}{105 } (\bibinfo{year}{2009}).

\end{thebibliography}

\end{document}